\documentclass[journal]{IEEEtran}
\usepackage{todonotes}
\usepackage{todonotes}

\ifCLASSINFOpdf
\else
\fi

\usepackage{hyperref}

\begin{document}
%



\title{NeuroRAN: Rethinking Virtualization for AI-native Radio Access Networks in 6G}

%
%
%

\author{Paris Carbone,~\IEEEmembership{Member,~ACM,}
        Gy\"{o}rgy D\'{a}n,~\IEEEmembership{Senior Member,~IEEE,}
        James Gross,~\IEEEmembership{Senior Member,~IEEE,}
        Bo G\"{o}ransson,~\IEEEmembership{Member,~IEEE,}
        Marina Petrova,~\IEEEmembership{Senior Member,~IEEE}

\thanks{P. Carbone, G. Dán, and J. Gross are with KTH Royal Institute of Technology, Stockholm, Sweden, E-mail: $\{parisc\vert gyuri\vert jamesgr\}$@kth.se}
\thanks{B. G\"{o}ransson is with Ericsson Standards \& Technology and KTH Royal Institute of Technology, Stockholm, Sweden, E-mail: bo.goransson@ericsson.com}
\thanks{M. Petrova is with RWTH Aachen University, Germany and KTH Royal Institute of Technology, Stockholm, Sweden, E-mail: petrovam@kth.se}
\thanks{Manuscript received April 15, 2021}}

%
%

\markboth{NeuroRAN}%
{Shell \MakeLowercase{\textit{et al.}}: Bare Demo of IEEEtran.cls for IEEE Journals}
%



\maketitle

\begin{abstract}
Network softwarization has revolutionized the architecture of cellular wireless networks. State-of-the-art container based virtual radio access networks (vRAN) provide enormous flexibility and reduced life cycle management costs, but they also come with prohibitive energy consumption. We argue that for future AI-native wireless networks to be flexible and energy efficient, there is a need for a new abstraction in network softwarization  that caters for neural network type of workloads and allows a large degree of service composability. 
In this paper we present the NeuroRAN architecture, which leverages stateful function as a user facing execution model, and is complemented with virtualized resources and decentralized resource management. We show that neural network based implementations of common transceiver functional blocks fit the proposed architecture, and we discuss key research challenges related to compilation and code generation, resource management, reliability and security.   
\end{abstract}


%
\IEEEpeerreviewmaketitle

\section{Introduction}
%
%
%
%
\IEEEPARstart{O}{ver} the last years, we are witnessing significant efforts in designing a software-based virtualized radio access network (vRAN) architecture running on commercial off the shelf (COTS) hardware, in an attempt to reduce development and maintenance costs, and to replace static and monolithic architectures with programmable and flexible ones. 
Software-based vRAN on COTS hardware indeed offers enormous flexibility. 
Flexibility is essential to support emerging applications that will tightly integrate with physical processes, e.g., augmented reality and cognitive assistants, real-time cyber-physical control systems, as well as situational awareness systems. 
At the same time it is a precondition for further antenna densificiation, which is the primary approach to providing increased capacity, lowering latency and increasing reliability. 
Yet, vRAN also comes with prohibitive energy consumption, and would require a tenfold increase in energy efficiency to allow wide scale adoption. Moreover, as of today, a fast and flexible programmable control framework, which can jointly meet real-time requirements of lower layers of the protocol stack, and can autonomously adapt to the time-varying network dynamics does not exist.

As 5G networks are rolled out all over the world, the requirements for next generation (6G) systems are just starting to be discussed in academia and in industry. 
The main application-level drivers for a future radio access network include increased trustworthiness and the ability to cope with emerging applications, such as XR/VR, gaming, smart sensors,  internet-of-sense applications, and digital twins. 
Many of these applications require or generate massive amounts of data and have tight delay requirements. 
At the same time, over the last few years we have been witnessing an unprecedented push in communications research towards data-driven approaches.
While the degree of functional involvement of machine learning varies so far with respect to the discussed approaches, many presented studies demonstrate on-par performance of data-driven approaches in comparison to legacy model-based ones.
Thus, it appears quite likely that next generation RANs will include widespread use of AI within transceivers.
At the same time, it goes without saying that they should be an enabler for a sustainable society, and that all this has to come at an affordable cost. 

\begin{figure*}[t]
\begin{minipage}[b]{1\linewidth}
    \centering
    \includegraphics[width=0.8\linewidth]{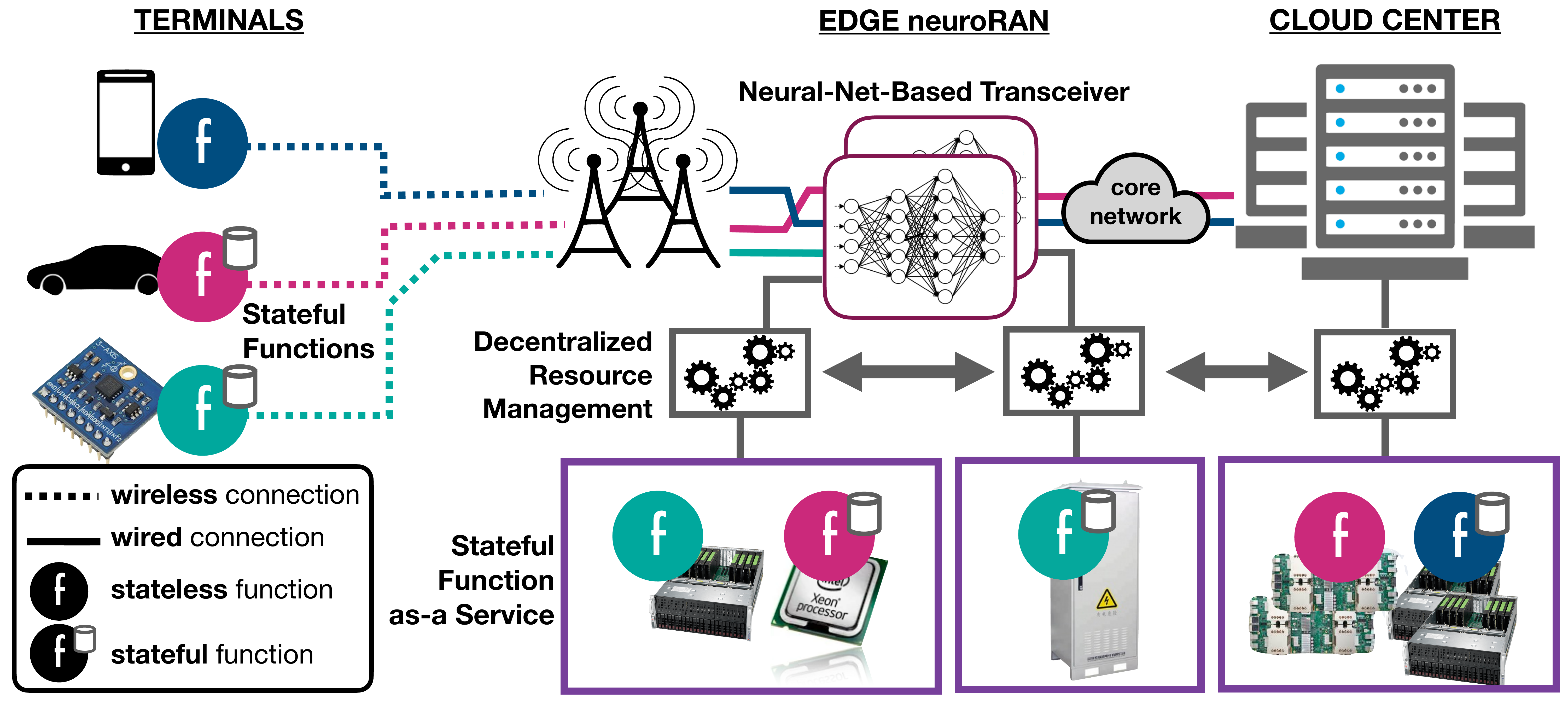}
    \caption{AI-native RAN Architecture Overview} 
    \label{fig:gcnsplit}
  \end{minipage} 
\end{figure*}

In this paper, we argue that the promises of ML in next-generation systems require suitable software architectures to actually deliver.
These go well beyond standard container-based approaches leveraged today with respect to flexible RAN solutions such as O-RAN \cite{O-RAN}.
The main components of such future architecture relate to efficient virtualization with respect to more adapt hardware, for instance, for neural network types of workloads, the provisioning of virtualized resources towards functional software blocks allowing fine-grained composition and function splits, as well as on-demand deployment of corresponding RAN transceiver code.
To this end, we present NeuroRAN, illustrated in Figure~\ref{fig:gcnsplit}, a matching software architecture for next-generation mobile networks, while also discussing newly arising challenges in the intersection between software architectures and ML-based transceiver software for future RANs. 

The reminder of the paper is organized as follows. In Section II we briefly discuss the evolution towards softwarized and virtualized mobile networks architectures and present current trends in deploying neural networks for design and adaptation of wireless communication in Section III. In Section IV we propose NeuroRAN, our energy-efficient and flexible AI-native RAN architecture and discuss the main research challenges to be addressed in Section V. Finally in Section VI we conclude the paper.

\section{Softwarization of Mobile Network Architectures}
Early product-grade LTE implementations were leveraging custom hardware and software in the RAN as well as in the EPC (evolved packet core). 
They were optimized with respect to energy consumption, but left little room for functional enhancements and required high coordination bandwidth among eNBs (base stations) with the advent of sophisticated interference coordination schemes.
RAN softwarization was arguably triggered by the emergence of software-defined networking (SDN) and network function virtualization (NFV) as new implementation paradigms for wide-area networks, allowing enhanced flexibility and separation of control and data plane functionality for enhanced control.
Vendors of LTE networks introduced the virtualized EPC, which too allowed a higher degree of flexibility and efficiency. 
Moreover, softwarization triggered the move away from the traditional non-split RAN architecture in LTE, i.e., where within the same cabinet all baseband processing and analog-to-digital processing is performed, to the split architecture in cloud RAN (C-RAN), where baseband units of different eNBs can be centrally pooled while A/D conversion is performed locally by radio units.
However, by being able to pool baseband units, in particular interference coordination was easily achieved in LTE Advanced by running associated and virtualized baseband units on the same resource pool.
Meanwhile, with the introduction of C-RAN the hardware dependency in the pool assignment has been lifted, allowing corresponding baseband units to be moved almost freely over a compute infrastructure.
Nevertheless, due to the legacy of the basic function definition of LTE systems, the potential of network function virtualization could not be fully leveraged.

The architecture of 5G systems has been defined from the start with softwarization and virtualization in mind.
Loosely speaking, the approach taken is to define smaller functional units that can be operated independently and thus be placed more flexibly over virtualized resources.
For the radio access network, this has led to the definition of radio units (RUs), distributed units (DUs) and central units (CUs) being defined as the building blocks of gNBs, the equivalent entities of LTE eNBs in 5G. 
The resulting architecture is more flexible, as only the RUs are dependent on specific hardware for A/D conversion and for analog handling of the signal.
In contrast, DUs and CUs can be executed on COTS hardware for realizing the lower and upper parts of the RAN network stack.
Frameworks like O-RAN or SD-RAN realize these vRAN implementations of 5G systems, and represent the state-of-the-art in fully softwarized mobile radio access networks. 

The downside to the increase in flexibility is increased power consumption.
While power consumption of a RAN results from the intimate relationship between architecture, implementation, run-time optimization and other aspects like cooling, focusing on the power consumption of a virtualized transceiver alone running on a general purpose processor steeply increases the consumption in comparison with processors custom-made for this purpose. 
This is not specific to any transceiver algorithm, but applies in general to almost all virtualized functions.
In existing wireless transceivers this has led to the use of customized hardware environments, mostly utilizing accelerators for certain functionality, at the price of loosing flexibility.
With the advent of massive MIMO and its exponentially increasing processing demands, the necessity to reduce the power consumption in future virtualized RANs becomes paramount.

\section{Neural Network Abstraction and Applications in Mobile and Wireless Networks}
The significant breakthroughs in neural networks for classification and pattern recognition associated with the advances in training acceleration through the use of GPUs about ten years ago, have led to enormous interest in using machine learning for various problems in communication systems.
In particular, deep learning (DL) has recently shown a great potential to become a powerful tool to design, optimize, adapt, and secure wireless communications. 

Deep learning makes use of deep neural networks (DNNs) which are cascades of parallel processing layers with individual connectivity degrees. 
In general, neural networks provide flexible abstractions to any functional input/output relationship at hand, for which sufficient training data is available. 
Through training, DL is capable of approximating functions and complex inter-relationships of variables that are hard to accurately describe using mathematical models. 
By doing so, DNNs enable novel approaches to the design of wireless communication systems without the knowledge of accurate mathematical models, e.g. unknown channel models.

While the potential performance improvements of a DNN-based wireless communication system design are currently receiving significant research attention, it is evident that flexibility of the implemented processing structure, as well as implementation costs and adaptability with the evolution of communication systems are strong additional arguments for the application of DNNs in communication systems.
In the following we will discuss few representative applications of DNNs to communications transceiver and protocol design.



\subsection{DNNs for Low-Layer Transceiver Architectures}
A plethora of research works have recently emerged studying machine learning applications to communication systems design. 
With respect to implemented architectures, works either focus on substituting individual functions in the transceiver chains, or more progressively substituting larger blocks, primarily in the physical layer.
Examples of the first category comprise for instance works on signal detection~\cite{Samuel_2019}, channel estimation~\cite{Neumann_2018}, or signal demapping in broadband wireless communication systems~\cite{Shental_2019}.
In all cases, it can be shown that deep learning, given sufficient training data, is either on par with legacy (model-based) approaches, or even outperforms them.
Depending on the application, this can come with a lower complexity of the learning approach. 

Works focusing on larger functional blocks typically propose to substitute several processing steps of the physical layer by a suitable aggregate DNN.
A good example is for instance the recent seminal work~\cite{Honkala_2021}. 
It introduces a DNN-based OFDM receiver implementation, converting frequency-domain signal samples into uncoded bits with soft information.
Thus, equalization, channel estimation and signal demapping are substituted by one trained neural network.
The work demonstrates the applicability of this approach to 5G new radio compliant signals, showing on par performance with legacy receiver structures.
Complexity-wise, the proposed neural network performs best with roughly 1 million parameters in a ResNet structure (a special type of connectivity structure of the neural network), while scaling it to larger bandwidths asymptotically becomes equivalent to LMMSE receivers.

Another fundamental approach to substituting entire blocks of transceivers is given by end-to-end approaches~\cite{Doerner_2018}.
Here, in contrast to~\cite{Honkala_2021}, the entire transmitter and receiver are substituted through DNNs, which allows for channel-specific signalling schemes.
In detail, variational autoencoders are utilized for the joint training of transmitter and receiver.
Training such special deep neural data structures encompasses an end-to-end consideration, giving the approach also its name.
In other words, end-to-end approaches are most consequent in moving away from model-based transceivers, potentially jeopardising traditional system standardization.
The authors showed that for narrowband, single-carrier systems the approach is in principle viable, achieving results that are on par with model-based implementations.
In terms of real-time performance, the implementation is not yet up to speed, though.

\subsection{DNNs for Higher-Layer Functionality}
While the bulk of information processing of any wireless transceiver is related to the physical layer, deep learning has been also considered higher up in the (wireless) network stack.
Focusing again first on works that consider to substitute individual functions of the network stack, efforts have been made for instance with respect to performing resource allocation for 5G networks by DNNs~\cite{Imtiaz_2021}.
Further works consider channel allocation for dynamic spectrum access~\cite{naparstek} or focus on improving sensing/classifying accuracy for dynamic spectrum access systems~\cite{kd}.

In contrast to substituting individual functions, a different category of works focus on automating the parameter tuning of communication protocols. 
When it comes to learning-based medium access protocol approaches, the research is in its infancy and mainly addresses learning optimal channel access policies. 
Dynamic protocol composition from a set of atomic components by means of deep neural networks has been presented in~\cite{Pasandi}.
More work has been done in the transport layer, where deep neural networks have been proposed to design congestion control algorithms and learn an optimal TCP congestion control policy from rich parameter observations of the network environment (e.g Queuing delay, inter-arrival times, round trip time (RTT), lost packets, etc.)~\cite{Zhang}. 
In contrast, conventional congestion control only considers several measurements such as packet loss and RTT as indicators of congestion, and cannot easily adapt to new networks or leverage experiences from the past. 

The emerging research works and the encouraging initial results on deploying DNNs in the design of communication system components and functions at different layers suggest an upcoming paradigm shift in the way how wireless and mobile networks will be architected in the 6G era and beyond. In the far future one could also envision that the future networks could be possibly designed by AI. Despite of this great promise, to pave the road towards native AI design a number of challenges will have to be resolved. Most notably, there will be a need for a software architecture that will support DNN processing and computation in a flexible but energy efficient way. In the following section we propose one such approach.

\section{Energy-efficient AI-native RAN: The NeuroRAN Architecture}

Neural networks show great promise for various baseband and PHY processing tasks, and could become building blocks of a future, flexible RAN architecture.
Yet, adopting neural processing on top of existing software abstractions will unlikely result in energy-efficient operation. Existing software and resource abstractions, virtual machines and containers, were optimized for data center environments with abundant, homogeneous hardware and for slowly changing environments. Consequently, they would result in significant memory and computational overhead in the implementation of an AI-native RAN. In lack of an abstraction for neural processing they would also fall short on flexibility due to the reliance on custom hardware accelerators, e.g., FPGAs, and cannot efficiently support adaptive composition of processing chains on short time scales. 

\begin{figure}[t]
\begin{minipage}[b]{1\linewidth}
    \centering
    \includegraphics[width=0.9\linewidth]{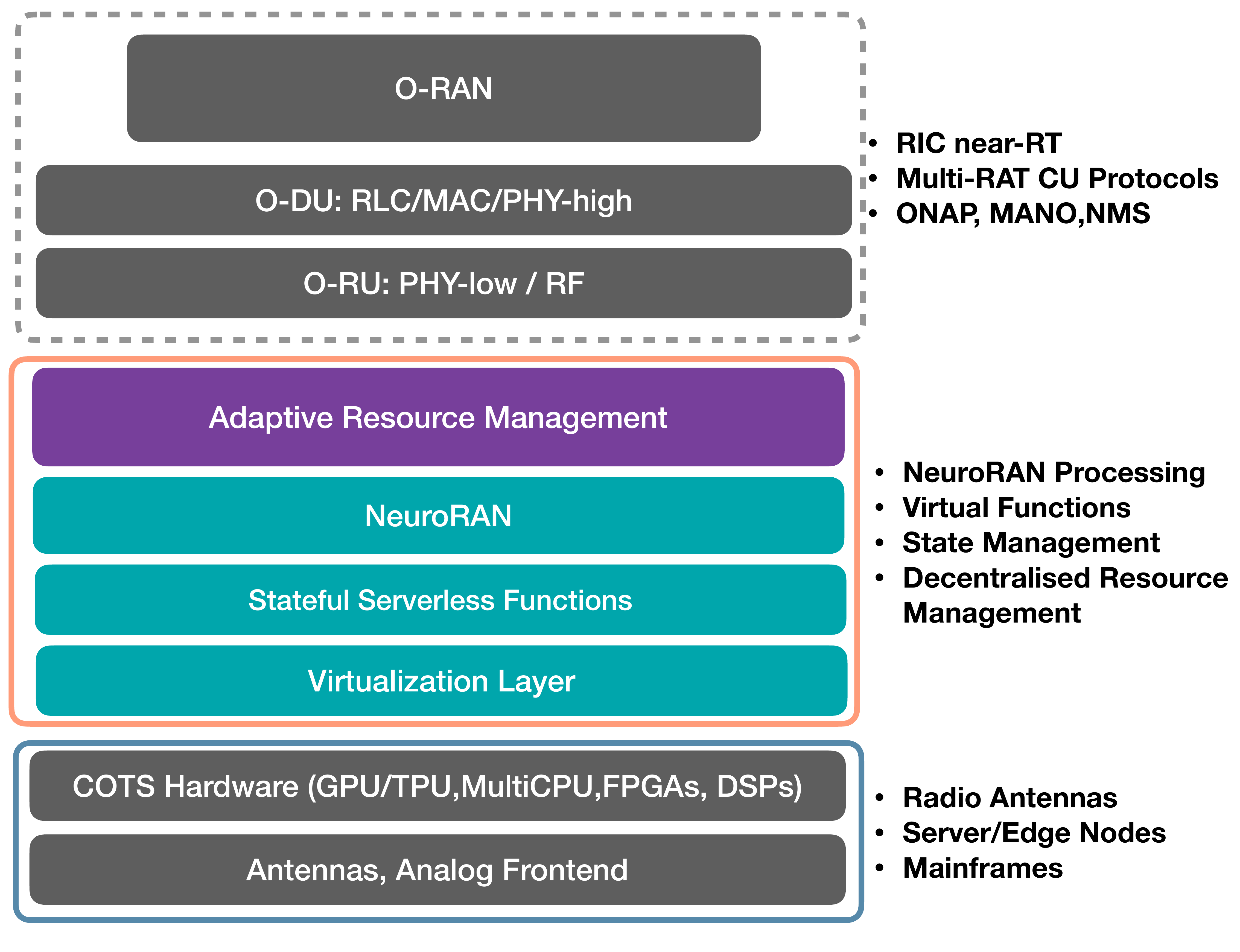}
    \caption{NeuroRAN Software Component Stack in the O-RAN Architecture.} 
    \label{fig:gcnsplit2}
  \end{minipage} 
\end{figure}

For a softwarized AI-native RAN to be energy-efficient and flexible, there is a need for a software architecture where AI-native functions are "first class citizens", as opposed to VMs and containers. 
The architecture should match the data-centric abstraction of computing that neural processing provides, i.e., agnostic to instructions sets, memory hierarchy, etc, which facilitates code reuse, portability and hardware maintenance. 
Furthermore, it should support fine-grained provisioning of resources, based on fine-grained compute actions without the need to allocate compute and network resources for long time periods (e.g., hours, days).

To address this gap, we propose the NeuroRAN architecture, which complements the O-RAN design through four main components, illustrated in Figure~\ref{fig:gcnsplit2}. 
First, we adopt the emerging paradigm of stateful function as a service (sFaaS) ~\cite{sreekanti13cloudburst}, which is becoming dominant in the domain of cloud virtualization for scalable applications~\cite{j2019berkeley}. 
Functions are ideal building blocks to compose decentralized dataflow applications (e.g., stateful streaming~\cite{carbone2017vldb}) and services, incorporating data subscription and session management and on-demand scalability (group communication, multiplayer gaming, collaborative editing apps, remote and VR control systems). 
Second, NeuroRAN,  a novel abstraction of neural-network on demand that can be supported and offered on top of stateful FaaS. NeuroRAN uses neural processing as a ubiquitous abstraction for common tensor-centric workloads extended with native support for tensor-centric hardware that is energy-efficient for tensor computations (e.g., FPGAs, TPUs) as a service.
Third, adoption of virtualization of compute, storage and communication resources to enable seamless migration of functions over heterogeneous hardware, alongside their corresponding state and data dependencies.
Fourth, decentralized resource management, which is itself a decentralized tensor-centric data flow application in the proposed framework, which can be migrated and executed on-demand. Table~\ref{tab::software-reqs} summarizes the main advantages of the proposed architecture compared to a state-of-the-art container-based architecture, as utilized for instance in O-RAN.

Starting from the bottom of the proposed middleware stack this leads to the following software components.

\noindent \textbf{RAN Virtualization Layer:} We propose the creation of a lightweight virtualization service that offers resource as a service capability to edge nodes with heterogeneous hardware in mind. This includes compute and memory resources such as GPUs, TPUs, flash memory, NVMe, RAM, and network resources. The virtualization layer should guarantee strong isolation and security of the resources, as well as the flexibility to compose and utilize custom configurations by combining memory, storage, network and compute components to facilitate the implementation of sFaaS on top. The virtualization layer should natively support the orchestration of the resources to meet performance metrics required by the utilized services (e.g., energy consumption, time allocation, local IO, network IO, vCPU instructions).

\begin{table}[t]
\centering
\caption{Feature comparison of container-based and stateful FaaS abstraction for AI-native RAN.}
\label{tab::software-reqs}
\begin{tabular}{c||c|c}
\hline
Software Feature & Container-based & sFaaS \\
\hline\hline
Memory footprint & Moderate & Low \\
\hline
CPU overhead & Significant & Low \\
\hline
Scalability & Custom & On-Demand \\
\hline
Maintenance & Custom & Automatic \\
\hline
Compute/Latency-Intensity & Custom & On-Demand \\
\hline
Hardware/Energy Consumption & Static & Dynamic \\
\hline
Composability & Coarse & Fine \\
\hline
Platform independence & No & Yes \\
\hline
Resource management & Coarse & Fine \\
\hline
Data path & Memory & Message passing\\
\hline
Reliability & Custom & Built-in \\
\hline
\end{tabular}
\end{table}

\noindent \textbf{Stateful Function as a Service:} As a user-facing execution model we propose the adoption of sFaaS. We envision the use of this paradigm both for lower network protocol functions (e.g., RU, DU, and CU) that exist in the O-RAN architecture as well as for the deployment of applications and services that can be executed on demand on top of edge nodes. The main differentiation to serverless functions known in cloud computing (e.g., Amazon Lambda) is the addition of explicit state and support for building end-to-end decentralized dataflow services composed out of interconnected functions that can communicate through remote invocations or via message passing.

\begin{table*}[!h]
    \centering
     \caption{Overview of Stateful and Stateless Transceiver Functional Blocks}
    \begin{tabular}{c|c|c|c}
    \hline
         Function & Stateful? & Parameters & State  \\
         \hline
         A/D Conversion & No & Baseband sampling frequency &\\
         STO, CFO correction & No & Signal type &\\
         PHY PDU extraction & Yes & & Scheduling grant\\
         Channel estimation & No & Signal type, pilot position &\\
         Equalization & No & Signal type, channel estimate &\\
         Signal (de)mapping & Yes & & Modulation scheme\\
         (De-)Interleaving & No & Interleaving scheme &\\
         Encoding, Decoding & Yes & & Coding scheme, coding block length\\
         CRC insertion,check & No & Block length, CRC scheme &\\
         En-/De-cryption & Yes & Key, encryption scheme & Cipher state\\
         \hline
    \end{tabular}
    \label{Table_Transceiver_State}
\end{table*}

\noindent \textbf{NeuroRAN:} As discussed in the previous section, most RAN functions that today are model-driven can be substituted by data-driven implementations. NeuroRAN allows for automating the deployment of data-driven implementations of RAN functions and other edge micro-services, including network functions (NFs) in the core network and emerging network data analytics function (NWDAF) services. NeuroRAN instances can be created, instantiated, trained and used for inference via a standard programming interface.

\noindent \textbf{Decentralized Resource Management:} A crucial component of the proposed middleware architecture is decentralized resource management. Supporting a FaaS deployment model requires the ability of the middleware to migrate functions and their corresponding state across nodes as well as scaling out resources elastically when needed. A resource management middleware can support and optimise these functions towards fair allocation of resources across decentralized edge nodes. Furthermore, this type of service has to be adaptive to cope with the dynamic nature of edge networks (failure detection, service discovery, reconfiguration etc.). In this setting service level agreements (SLAs) can specify requirements in terms of latency, throughput, and availability, and the objective of resource management is to minimize energy consumption subject to meeting all SLAs.

While many candidate application workloads for edge computing are inherently tensor-centric, such as visual analytics, autonomous industrial systems, cars and drones, a fundamental question for the feasibility of the proposed architecture is whether typical transceiver functional blocks can be implemented in the FaaS model. Table~\ref{Table_Transceiver_State} shows a list of radio transceiver functional blocks, their typical parameters and whether they require state to be maintained. The table indicates that out of ten functional blocks four require a stateful FaaS implementation, and for three of those the state is due to their dependence on scheduling decisions, which are performed at the same time scale as the functions would be invoked.

\section{Research Challenges}
The proposed architecture introduces a set of core challenges that we foresee are necessary to be addressed. We categorize them in a bottom-up fashion from the virtualization layer up to resource and service management.

\noindent \textbf{Virtualization over Heterogeneous Hardware:} 
At the lowest level we identify a set of problems that need to be addressed in order to enable seamless support for sFaaS, ranging from compilation and code generation to deployment and provisioning. Heterogeneous hardware should be supported out of the box without the need for reconfiguring and re-compiling libraries in edge nodes. To that end, there is a need for on-demand code generation techniques that can translate high-level sFaaS program specifications to low-level instructions supported by the underlying hardware (GPUs, multicore x86, TPUs and DSPs, for compatibility with legacy RAN hardware). Recent works in compiler research adopt intermediate code representations and build on widely adopted LLVM libraries, such as multi-level intermediate representation (MLIR~\cite{lattner2020mlir}), which already support a number of existing  representations and compilation tools for current and upcoming hardware architectures. Finally, we foresee it to be necessary to employ decentralized provisioning of services using sFaaS with respect to 3 core usage metrics: aggregated number of invocations, IO (state size) and data transferred across functions. The management of all the provisioning of these metrics needs to be made consistently and efficiently, which is a challenging task itself in a heterogeneous environment.

\noindent  \textbf{Resource Management in Dynamic and Heterogeneous Environments:}
A key enabler of the proposed architecture is decentralized scheduling enabling on-demand resource allocation, instantiation, migration and invocation of stateful FaaS. Existing approaches to resource management do not apply well to neural network abstractions and cannot provide throughput and latency guarantees in highly distributed environments. Recent efforts on applying deep learning to resource management would need to be extended to the multi-agent setting, but the convergence and stability of the resulting systems is not well understood.  A promising direction could be the use of graph convolutional network (GCN) embeddings, possibly starting from an initial model through iterative refinement and including performance SLAs as multi-objective optimisation variables in the training process of the network. Doing so could provide predictable performance in the spirit of safe machine learning. 

\noindent \textbf{Reliability and Security:} 
Among the biggest challenges of adapting cloud computing technologies to decentralized edge networks are reliability and security. Reliability not only involves provisioning of resources in response to failures, but also processing guarantees (number of times a function is executed) and consistency guarantees (for service state) despite failures. This is especially challenging in edge networks under high churn. In addition, FaaS requires strong isolation guarantees (process memory, non-volatile storage, virtual CPUs) offered by the underlying virtualization layer, which is important both for accurate provisioning as well as for the secure execution of FaaS. Finally, state management for FaaS over COTS opens many new challenges such as the need for automated state partitioning, migration, encryption-support as well as consistency guarantees at a dataflow- or FaaS-level in the presence of partial failures.


\section{Conclusion}
Motivated by the increasing importance of energy-efficiency in RAN operations, in this article we discussed the requirements that the migration to an AI-native RAN would impose on software abstractions in beyond 5G networks. We presented arguments that show that combining flexibility with energy efficiency would require going beyond existing abstractions, and could be possible by extending the emerging serverless computing paradigm to a stateful FaaS model combined with a neural abstraction of computing functions. The proposed architecture has the potential to meet the performance and reliability requirements of beyond 5G wireless networks, but it remains to understand what tradeoffs it involves in terms of security.


%



\ifCLASSOPTIONcaptionsoff
  \newpage
\fi

\end{document}